\documentclass[5p]{elsarticle}

\usepackage{graphicx}
\usepackage{amssymb}
\usepackage{amsmath}
\usepackage[section]{placeins}
\usepackage{etoolbox,lineno}
\usepackage{textcomp}
\usepackage{hyperref}

\hypersetup{colorlinks=true,linkcolor=blue, linktocpage}

\setcitestyle{authoryear}
\setcitestyle{round}
\setcitestyle{semicolon}

\makeatletter
\def\ps@pprintTitle{%
 \let\@oddhead\@empty
 \let\@evenhead\@empty
 \def\@oddfoot{}%
 \let\@evenfoot\@oddfoot}
\makeatother

\begin{document}

\let\today\relax

\begin{frontmatter}

\title{Lagged correlation-based deep learning for directional trend change prediction in financial time series}

\author[first]{Ben Moews\corref{corresponding}}
\ead{b.moews@ed.ac.uk}
\cortext[corresponding]{Corresponding author}

\author[second]{J. Michael Herrmann}
\ead{michael.herrmann@ed.ac.uk}
\author[third]{Gbenga Ibikunle}
\ead{gbenga.ibikunle@ed.ac.uk}

\address[first]{School of Physics \& Astronomy, University of Edinburgh, Edinburgh, EH9 3HJ, UK}
\address[second]{School of Informatics, University of Edinburgh, Edinburgh, EH8 9AB, UK}
\address[third]{Business School, University of Edinburgh, Edinburgh, EH8 9JS, UK}

\nonumnote{\textcopyright 2018. This manuscript version is made available under the CC-BY-NC-ND 4.0 license (\url{http://creativecommons.org/licenses/by-nc-nd/4.0/})\\}

\begin{abstract}
Trend change prediction in complex systems with a large number of noisy time series is a problem with many applications for real-world phenomena, with stock markets as a notoriously difficult to predict example of such systems. We approach predictions of directional trend changes via complex lagged correlations between them, excluding any information about the target series from the respective inputs to achieve predictions purely based on such correlations with other series. We propose the use of deep neural networks that employ step-wise linear regressions with exponential smoothing in the preparatory feature engineering for this task, with regression slopes as trend strength indicators for a given time interval. We apply this method to historical stock market data from 2011 to 2016 as a use case example of lagged correlations between large numbers of time series that are heavily influenced by externally arising new information as a random factor. The results demonstrate the viability of the proposed approach, with state-of-the-art accuracies and accounting for the statistical significance of the results for additional validation, as well as important implications for modern financial economics.
\end{abstract}

\begin{keyword}
Lagged correlation \sep Deep learning \sep Trend analysis \sep Stock market
\MSC[2010] 68T05\sep 62P20
\end{keyword}

\end{frontmatter}

\nolinenumbers

\section{Introduction}
\label{S:1}

An increased interest in deep-layered machine learning approaches for time series analysis and forecasting resulted in applications in various fields, establishing this area as a challenging topic of interest \citep{Cao2003Support, Nesreen2010An}. When it comes to the effective use of deep neural networks, one of the primary concerns is a sensible approach to feature engineering for useful data representations. This process often depends on domain knowledge about the respective area of application and is, more often than not, a time-consuming part of research \citep{Najafabadi2015Deep}. Some researchers equate applied machine learning, in an attempt to emphasize the relative importance, with the concept of feature engineering itself \citep{Ng2012Machine}. Such representations have to be informationally rich enough to incorporate the looked-for lagged correlations between time series while, at the same time, being constrained to a discrete number per observation and variable for input features in a feed-forward neural network. \citet{Zhang2005Neural} find that such models are not able to capture the necessary information when applied to raw data from time series with seasonal and trend patterns, which opens the field for approaches to feature engineering that allow for an effective use of time series data for trend predictions in a variety of application areas.

In this paper, we test the hypothesis that deep feed-forward neural networks, combined with exponential smoothing for the training inputs, are suitable for learning lagged correlations between the step-wise trends of a large number of time series, and that such models can be successfully applied to current research on real-world forecasting problems. In order to test this approach, we apply the proposed method to gradients computed for five years of historical stock price data of the S\&P 500 stocks in one-hour intervals for daily trends, adding the complication of relatively few observations. For a more in-depth overview of soft computing methods in financial market research, interested readers are referred to \citet{Cavalcante2016Computational}, with \citet{Weng2018Predicting} providing an application of ensemble methods to financial markets using a variety of text-based and index-based features.

The experiments that are conducted for this purpose demonstrate the viability of our approach by predicting price trend changes with an accuracy above given market baselines and within a stringent statistical validation framework. In order to evaluate the soundness of our conclusions, we test the results against the alternative possibilities of simply learning frequencies or probabilistic distributions, calculate confidence intervals and \textit{p}-values, and perform a visual analysis via notched box plots \citep{McGill1978Variations}. The results of this paper deliver evidence for the applicability to a number of real-world problems that deal with complex relationships of large numbers of noisy time series.

Our hypothesis relates to the microstructure model by \citet{Ho1983The}, which characterizes the links between quote changes in a stock and the evolution of the inventory with respect to the other stocks. The model shows that quote changes in stock $a$, which is in reaction to a transaction in stock $b$, are based on $\textup{cov}(R_a, R_b) / \sigma^2(R_b)$. This portfolio view of stock trading is in line with commonly deployed diversification strategies in finance and has given rise to highly popular instruments such as exchange traded funds (ETFs), which offer cheap means of diversifying risk. This study has implications for perhaps the longest running debate in the financial economics literature.

The unpredictability of the factors influencing price discovery in stocks makes the price discovery process noisy \citep{Chen1986Economic}. The unpredictability (or randomness) of the information acquisition process in financial markets is consistent with the efficient market hypothesis described by \citet{Fama1965The} and \citet{Fama1970Efficient} as well as the random walk hypothesis \citep{Kendall1953The, Cootner1964The, Malkiel1973A}. These theories contradict our hypothesis on the existence of time-shifted correlations in stock markets. Consistent with our expectations, our results deliver rigorously tested empirical evidence, supporting the existence of time-shifted correlations in stock prices and thus contradict the random walk theory. Specifically, our findings are inconsistent with \citet{Sitte2002Neural}, who argue that the price discovery process for S\&P 500 stocks is a random walk due to the inability of artificial neural networks to extract any information resulting in above-average predictions for those stocks.

Our results are, however, consistent with previous, albeit weak, evidence for the absence of a random walk in financial time series via the use of artificial neural networks as presented by \citet{Darrat2000On}. The consistency of the efficient market hypothesis with the random walk hypothesis also implies that our findings contradict much of the efficient market hypothesis, which is widely supported by a large section of the finance academic literature \citep{Fama1970Efficient, Doran2010Confidence}. However, despite the seemingly established nature of the random walk hypothesis, many studies, like \citet{Lo1987Stock}, have questioned its validity over the years, while others have proposed alternatives. For example, one popular alternative hypothesis is that stock returns may be explained by the sum of a random walk and a stationary mean-reverting component \citep{Summers1986Does, Fama1988Permanent}. \citet{Lo1987Stock} also advance the view that the efficient market hypothesis is an `incomplete hypothesis'. This current paper is not an attempt to reconcile one of the longest-standing debates in the finance literature; rather, we propose a price prediction approach based on an amalgamation of market microstructure theory and machine learning.

\section{Related research}
\label{S:2}

\subsection{Trend prediction in time series}
\label{S:2.1}

The feasibility of different types of artificial neural networks for trend prediction in time series was indicated early by \citet{Saad1998Comparative}. Other types of noise reduction, for example PCA for echo state networks, have already been subjected to similar investigations, with no significant success being reported \citep{Lin2009Short}. While research on regression gradients as input features for feed-forward neural networks is sparse in the published literature, the usage of directional derivatives of wavelets was recently introduced in natural language processing \citep{Gibson2013Spectro}.

Features based on derivatives were subsequently adapted in other research areas, for example statistics and digital signal processing \citep{Gorecki2013Using, Baggenstoss2015Derivative}. The success of the regression derivative-based approach for trend forecasting using lagged correlations between time series presented here provides additional evidence for the viability of such methods for time series applications. Specifically, positive results show the value and applicability of deep learning methods for such scenarios.

\subsection{Stock markets as a use case}
\label{S:2.2}

Price changes in stock markets are, at their core, the result of human decisions which, in turn, are based on their respective beliefs about stocks' future performance. Stocks are influenced not only by a company's respective performance, but also by newly arising information not directly linked to the latter. Examples are negative effects on stock prices for airlines after the 9/11 attacks, and similar effects after news about a CEO's diminishing personal health \citep{Drakos2004Terrorism, Perryman2010When}. Price changes are, therefore, the result of human beliefs about the future beliefs of other humans, which can be iterated indefinitely and is an example of real-world time series being created by human decision-making as well as the implementation of automated decision-making based on these notions, especially in high frequency trading.

It can be concluded that time series of historical stock prices contain, due to these factors, a large amount of noise in the form of new information influencing the process, and through biases and errors in human decision-making. Markets are, as a result, inherently prone to fluctuations triggered by overreactions, and to dynamical reinforcement during temporary crazes \citep{Chen1986Economic}. This makes them, due to the complexity of their generation process, a suitable use case to test the proposed model's ability to identify and exploit lagged correlations in notoriously hard-to-predict noisy environments.

Relevant recent work on stock market prediction includes \citet{Zhang2009Stock} on changes to backpropagation, \citet{Boyacioglu2010An} on the use of fuzzy inference frameworks, \citet{Chatzis2018Forecasting} on deep learning for financial crisis forecasting, \citet{Zhang2018A} on unsupervised heurstic algorithms, and \citet{Malagrino2018Forecasting} on Bayesian network approaches. Another area of research is concerned with text-based stock market prediction, with \citet{Nassirtoussi2014eText} providing a holistic overview for interested readers.

\subsection{ANNs and stock markets}
\label{S:2.3}

The viability of using artificial neural networks for stock market predictions was first hypothesized by \citet{White1988Economic}, with subsequent indications of success by \citet{Saad1998Comparative} and \citet{Skabar2002Neural}. \citet{Zhang1997Forecasting} report on the special suitability of artificial neural networks to such forecasting problems due to their adaptability, non-linearity and arbitrary function mapping. \citet{Takeuchi2013Applying} first made experimental results on the use of deep neural network models for stock market prediction available as a working paper by using the work of \citet{Hinton2006Reducing}, and with a binary prediction accuracy of 53.36\%. A similar approach by \citet{Batres-Estrada2015Deep} resulted in a comparable reported accuracy of 52.89\%, also outperforming a simple logistic regression.

\section{Gradients as features}
\label{S:3}

\subsection{Approximating trend strengths}
\label{S:3.1}

Linear regressions are a wide-spread approach to capture trends limited to a certain time frame and, as such, form the basis for these features. They take, in their general form, the following shape, with $i$ indicating one of $m$ observations, intercept $\beta_0$ and error term $\epsilon_i$:

\begin{equation}
\begin{aligned}
y_i \ &= \ \beta_0 + \beta_1 x_{i, 1} + \beta_2 x_{i, 2} + \ , \ ... \ , + \beta_p x_{i, p} + \epsilon_i \\
 &= \ \mathbf{x}^T_i \boldsymbol{\beta} + \epsilon_i \ , \ \ i \in \{ 1 \ , \ ... \ , \ m \}
\end{aligned}
\smallskip
\end{equation}

As a one-variable feature is necessary for the model's input vector, a simple linear regression is used as a least-squares estimator of equation (1). This estimator identifies one explanatory variable to minimize the squared sum of the residuals by fitting a line. The equation is presented as a minimization problem seeking $\beta_0$ and the slope $\beta_1$ for $\underset{\beta_0, \beta_1}{min} \ Q(\beta_0, \beta_1)$:

\begin{equation}
 Q(\beta_0, \beta_1) \ = \ \sum_{i=1}^m (y_i - \beta_0 - \beta_1 x_i)^2
\smallskip
\end{equation}

The gradient of the resulting regression model, that is, the slope of the fitted line, can then be computed by taking the first derivative. In the given task, this means extracting $\beta_1$ as a feature. Information about the value of a time series at a point along the timeline is lost in this process, with the resulting gradient representing the strength of an upward or downward movement of a trend via the regression model. A time interval size over which the regression is to be performed has to be determined in advance. The feature matrix, which is later used for the input of a feed-forward neural network, consists of the derivatives w.r.t. $\beta_{1,k}$ of a simple linear regression as in equation (2) per time series $k \in S := \{1 \ , \ ... \ , \ s \}$, and for all separate time intervals $j \in T := \{1 \ , \ ... \ , t \}$, meaning that $\forall \ j \in T, \ k \in S$:

\begin{equation}
\begin{aligned}
& \frac{\partial}{\partial \beta_{1,k}^j} \ \ \underset{\beta_{0,1}^j, \beta_{1,1}^j}{min} \ \sum_{i=1}^m (y_{i,k} - \beta_{0,k}^j - \beta_{1,k}^j x_{i,k}^j)^2 \\ \\
& \Rightarrow
\left(
\begin{tabular}{cccccc}
$\beta_{1,1}^1$ & $\beta_{1,1}^2$ & . & . & . & $\beta_{1,1}^t$ \\
$\beta_{1,2}^1$ & $\beta_{1,2}^2$ & . & . & . & $\beta_{1,2}^t$ \\
. & . & . & & & .\\
. & . & & . & & . \\
. & . & & & . & . \\
$\beta_{1,s}^1$ & $\beta_{1,s}^2$ & . & . & . & $\beta_{1,s}^t$
\end{tabular}
\right)
=: \mathbf{B}
\end{aligned}
\smallskip
\end{equation}

When applied to all time intervals, this resulting feature matrix $\mathbf{B}$ of respective gradients contains directional trend strength indicators for all time series, one per row, and with one time interval per column. This representation of time interval trends is the basis for the subsequent smoothing process.

One point of concern is the reduction of the dataset due to the given time interval over which a trend is approximated, which requires a sufficiently large set of observations to effectively train artificial neural networks after the reduction \citep{Glorot2010Understanding}. Research on the usage of regression derivatives for predictive classification tasks with time series, especially with regard to deep learning approaches, remains sparse. Examples from recent years include the use of such gradients for audio classification via decision trees and support vector machines by \citet{Mierswa2005Automatic}, gradients of wavelets for tasks in natural language processing by \citet{Gibson2013Spectro} and the addition of such derivatives to dynamic time warping \citep{Gorecki2013Using}.

\subsection{Infinite impulse response filtering}
\label{S:3.2}

\citet{Clarke2000The} and \citet{Gehrig2006Extended} state that technical analysis, that is, the prediction of stock price changes based on historical stock market data in the form of time series, is wide-spread in the current investment industry. The exponential moving average is, in this context, one of the dominant techniques used as a lagged indicator for technical analysis. In the general study of time series analysis, it is better known as exponential smoothing \citep{Brown1956Exponential, Holt1957Forecasting}. It serves as a type of infinite impulse response filtering that can be computed recursively as follows, with $\alpha$ being the smoothing factor used, with $\alpha \in (0, 1)$ and $t$ as the time interval indicator:

\begin{equation}
\begin{aligned}
s_0 & = x_0 \\
s_t & = \alpha x_t + (1 - \alpha) s_{t-1} \ , \ \ t \ > \ 0
\end{aligned}
\smallskip
\end{equation}

Exponential smoothing is a staple method in digital signal processing, and special consideration has to be placed on the choice of the smoothing factor $\alpha$. The sensitivity of a prediction w.r.t. $s_0$ is inversely correlated with the size of $\alpha$, with the average of at least 10 time intervals being widely recommended as the initial value for $s_0$ \citep{Nahmias2009Production}. The applicability of infinite impulse response filters to the smoothing of financial time series has first been emphasized by \citet{Gencay2001An}.

\section{Empirical validation}
\label{S:4}

\subsection{Data cleansing and pre-processing}
\label{S:4.1}

The dataset, which is provided by Thomson Reuters Tick History, features the average stock price per hourly time step from 2011-04-04 to 2016-04-01, covering about five years worth of stock market information for the current 505 S\&P 500 stocks, with a combined number of 6,049,849 observations. It contains the price, date and time of an observation and the respective stock's Reuters Instrument Code (RIC).

For an effective use as feature vectors, and to avoid a contamination from a financial perspective, the price series over which the step-wise gradients are computed have to be perfectly aligned and existent for all time steps and used stocks. As the dataset is imperfect, invalid time stamps and non-consistent values for holidays are sorted out, and missing values are reconstructed via the next preceding value of a row with the same RIC. The RICs for which a cut-off value for a minimal number of existent observations is present are then kept. The rest of the stocks, 56 in total, are discarded, as a further reconstruction process would endanger the dataset of becoming corrupted through too many reconstructed insertions.

To obtain a perfect alignment despite missing rows, we devised an algorithm optimized for speed that operates on two combined date-time vectors; one from an uninterrupted timeline and one per stock matrix, with the dataset being split into a list of matrices with one list place per RIC.

For each list place, the matrix is inflated to avoid higher computational costs via appending rows. At each discrepancy between the ideal and actual time-date vectors, the original matrix within the inflated matrix is shifted one place down, the missing row is substituted with the next adjacent values, and the missing date-time stamp is inserted. As this process operates on the fine-grained level of the raw data and only replaces very few missing values, for example for initial public offerings after the stock market opens and earlier market closings on Thanksgiving, the result on trend features over larger time intervals is negligible.

At the end, each matrix is cut at the first occurrence of the last date-time stamp, and only price information per stock is retained. With this information, price trend approximations are computed, for each trading day and stock, with the simple linear regression from equation (2). This leaves us with 1,241 observations for 449 stocks for feature matrix \textbf{B} from equation (3), on which the exponential smoothing from equation (4) is performed.

\subsection{Setup of the use case experiments}
\label{S:4.2}

The experimental setup shown in Figure 1 ensures that the experiments test for time-shifted correlations between time series instead of using a stock's own historical information, that is, data of the predicted stock is not part of the model's input. This is one of the main differences to related work on time series-based stock market prediction, for example \citet{Takeuchi2013Applying}, with the accuracy used as the metric by which the presence of correlations is measured. Another difference is the use of more short-term, meaning daily instead of monthly, predictions. As we aim to find time-invariant correlations between stocks, five-fold cross-validation with a validation set for early stopping is used to reduce the variability of results, and to use data in a frugal manner \citep{Seni2010Ensemble}.

\begin{figure}[!h]
\includegraphics[width=\columnwidth]{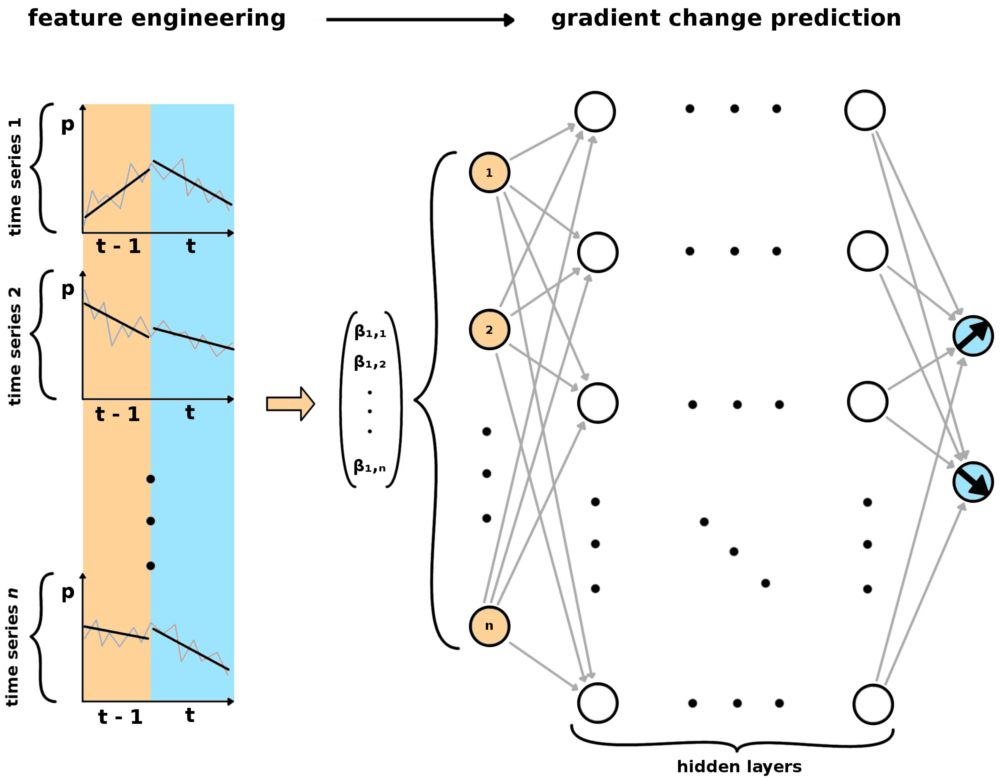}
\caption{Schematic depiction of the gradient calculation and deep classification model process. For each time series, each time interval's trend approximation is computed via a simple linear regression over a pre-determined time window, the gradient of which is then extracted to form a time interval's gradient vector $(\beta_{1, 1}, \beta_{1, 2}, \ ... \ , \beta_{1, n})$ as one column of the feature matrix. The latter is then used to train a deep feed-forward neural network for the binary prediction of directional trend changes.}
\label{fig:1}
\end{figure}

This way, a 60-20-20 split for training, validation and test sets is used for each fold and stock. When performing \textit{k}-fold cross-validation, a central concern is the choice of the correct split ratios. The 60-20-20 split is a frequently used rule of thumb in machine learning in which a validation set, in addition to the training and test sets, are necessary. Regarding the larger size of the training set, we want to ensure a large-enough number of data points in the training set to enable the model to learn sufficiently. Conversely, if the validation set is too small, the validation of the learned parameters cannot be properly assessed with a representative sample, while an insufficient size of the test set leads to the final test on unseen data being similarly non-representative.The training examples split from the dataset are normalized element-wise using min-max scaling. Exponential smoothing, as described before, is then applied to the input features of all three sets.

We repeat these experiments for each stock as the respective target, without past information concerning the stock itself in the inputs. The output vector takes the form of a one-hot representation to indicate whether gradients are larger or smaller than for the preceding interval. Two output nodes are chosen instead of one in accordance with the results of \citet{Takeuchi2013Applying} and \citet{Ding2015Deep} in favor of this setup, and to ease the structural comparability.

Preliminary test runs, with 20 randomly chosen stocks to avoid cherry-picking hyperparameters for the use case in general, showed the smallest test set error for 400 nodes per hidden layer, as measured in increments of 50 nodes for up to 800 nodes. The final model used for the experiments features 10 hidden layers, a number that was determined by starting with one hidden layer and, subsequently, adding $4 - n$ hidden layers for $n \in \{0, 1, 2, 3\}$, with the option to add further single layers in case of still-increasing performance. With improved performance up to the addition of nine additional hidden layers, the performance plateaued at this point. Early stopping, together with $\ell_2$ regularization, is used to prevent overfitting and unnecessary complexity, whereas momentum is applied to prevent stochastic gradient descent from terminating in small-spaced local minima. Dropout, another popular method of preventing overfitting in neural network, is a popular technique in deep learning \citep{Hinton2012Improving, Srivastava2014Dropout}. Initial experiments with different levels of omissions did, however, prevent the model from learning, which can be ascribed to the very high levels of noise in stock market information. Epoch-based learning rate decay is used to find a minimum along the optimizer's descent path, with learning rate $\nu$, decay coefficient $\delta$ and epoch number $e$:

\begin{equation}
\mu_e \ = \ \mu_{e - 1} \ \cdot \ \frac{1}{1 + \delta \ \cdot \ e}
\smallskip
\end{equation}

Hyperbolic tangent functions serve as activation functions, with sigmoid functions at the output layer. The former choice is due to the intent to combat weight saturation, while the latter function is chosen over the softmax function due to the interpretability of the results as independent probabilities, and because these results are not needed to integrate to 1 as inputs for subsequent methods.

The model's weights are initialized as scaled samples from a zero-mean Gaussian distribution to address the potential of vanishing or exploding gradients, with a variance of $\frac{2}{n_l}$ and an initial bias of 0, and with $n_l$ denoting the number of connections in the $n$-th layer, allowing for an easy adaptation to future experiments with rectified linear units \citep{Glorot2010Understanding, He2015Delving}.

\subsection{Accuracy and validation measures}
\label{S:4.3}

For each stock and fold in each model, a randomly shuffled copy of the predictions is created and tested against the correct targets in addition to the predictions, resulting in mock predictions with a distribution identical to the actual predictions. This copy can be used to test whether the model just learns the distribution of the two output classes in the training set, which would result in very similar accuracies for the actual and mock predictions when compared to the correct targets.

Another case that has to be ruled out is that of a model learning to predict the dominant class of a training set. Two targets are created for each stock and model, each containing only one of the two classes. A model that learns more actionable information from its respective inputs than the dominant class of the training set needs to perform better on the test set than both these one-class mock targets in direct comparison to the correct targets. A fourth validation metric is computed by taking the maximum of all three metrics' accuracies for each time step, ensuring that a baseline with at least 50\% accuracy in each time step is reached in case of distributions that deviate from a 50:50 distribution. As the use of accuracy as the metric of choice can lead to questionable results, we randomly samples 50 of the dataset's stocks and evaluates them for both accuracy and the area under the curve (AUC) for the receiver operating characteristic, with the latter being a well-established metric for skewed class distributions \citep{Bradley1997The}. While the classes in our work are well-balanced, this exercise provides a validation of the choice of metric, and shines a light on the necessity of taking class distributions into consideration when dealing with classification problems. The accuracy for this subset of stocks is 55.284\%, while the AUC score is 0.55278. The necessity of using an additional decimal number demonstrates the similarity of both metrics for well-balanced classes such as the ones used in this paper. Additionally, a linear support vector classifier (SVC) is used on the same gradient-based data to enable a comparison with a simpler approach.

The average accuracy over all stocks is given as the standard method to assess the predictive power of a model. These accuracies do, however, not indicate whether the predictions' variations are too large to be considered successful, that is, whether the volatility of the model grows too large. For this reason, we also provide accuracies for the three baseline mock predictions. In addition, the \textit{p}-values for the predictions' accuracies via an upper-tail test are calculated for each of the baselines and an additional baseline that contains the highest accuracy among the three baselines for each stock, that is, for each model. The null hypothesis $H_0$ in each case is that the predictions' accuracies are not significantly larger than the respective baseline, with a very strict significance level of $\alpha \ =$ 0.001.

Lastly, notched box plots are a commonly used visualization tool for descriptive statistics, using the respective data's quartiles to allow for an intuitive representation. Non-overlapping notches for two boxes indicate a statistically significant median difference at 95\% confidence. Welch's \textit{t}-test is used to achieve higher reliability for unequal variances.

In order to rule out the simpler explanation that only a few observations are sufficient to identify general market behavior, and to address the possibility that using only past data to predict future trend changes yields no better accuracies than cross-validation that also takes future observations into account, an additional experiment is conducted: For 20 randomly sampled non-repeating stocks, and with $N$ being the size of the whole dataset, a neural network model is first fully trained, as for the primary experiments, on the first observation. Then, for $l = 2$ to $(N-1)$, the model is iteratively updated by training for 3 epochs on observations 1 to $l$ as the training set, while being tested on observations $l+1$ to $N$ as the test set. This procedure can be visualized as sliding a divisive line along across the dataset, training on an ever-increasing training set while simultaneously reducing the test set. The last point has to be kept in mind, as a small test set does not offer a good representation of the data. For this reason, the described procedure is stopped at the point when only 10\% of observations remain in the test set to still deliver a viable estimate.

\subsection{Results of the experiments}
\label{S:4.4}

The use of \textit{p}-values has to be viewed with caution, as criticism of their often incorrect use has risen in recent years. In 2016, the American Statistical Association published an official warning regarding the wide-spread misuse of \textit{p}-values \citep{Wasserstein2016The}. Accordingly, the \textit{p}-values are given in combination with other metrics such as the lower boundary for differences in means given a 99.9\% confidence interval, notched box plots for median differences and quartile distributions, and accuracies for models and baselines.

Figure 2 shows the accuracy of 58.10\% listed in Table 1 significantly above all baselines, both for the means as measured by the \textit{p}-values and the medians as indicated by the box plots, with neither the notches nor the boxes overlapping. The first and third quartiles are, however, spread wider for the model when compared to the baselines, with the exception of the SVC. In Table 1, the accuracies for the model, the different baselines and a simple linear support vector classifier are given, as well as the \textit{p}-value results with regard to the means and the minimal difference for a 99.9\% confidence interval.
 
\begin{figure}[!h]
\includegraphics[width=\columnwidth]{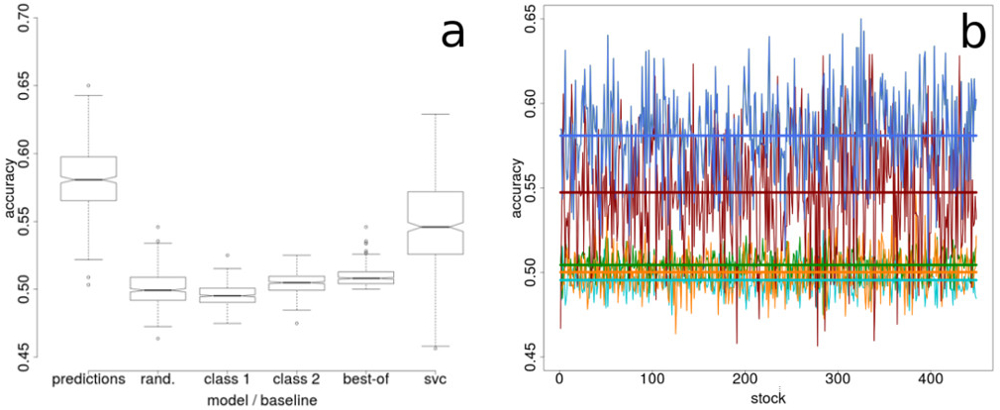}
\caption{Statistical validation and accuracies. Part $a$ shows a notched box plot representation of the results. Non-overlapping notches indicate a statistically significant difference in medians at 95\% confidence, which is the case for all baselines. Here, \textit{class 1} is the forecast that stock trends will change downwards, while \textit{class 2} refers to an upward change. \textit{predictions} denotes the performance of the model\textit{, svc} shows the boxplot for the support vector classifier (SCV), \textit{random} is the performance of a random forecast with the same distribution as the actual predictions, and \textit{best-of} represents the best result among the baselines for each time step. Part $b$ shows an intuitive visualization of the mean accuracies. The blue line represents the model's predictions and the red line the SVC's predictions, whereas predicting solely upward changes or downward changes is drawn in light blue and green, respectively. Fulvous denotes the random vector with the same class distribution as the model's predictions. The averages for the model results and the randomized predictions are drawn as lines of the same color.}
\label{fig:2}
\end{figure}

\begin{table}[!h]
\centering
\begin{footnotesize}
\label{tab:1}
\begin{tabular}{l l l l l l}
\hline
\multicolumn{6}{l}{\textbf{accuracies of predictions}} \\
\hline\noalign{\smallskip}
model & rand. & class 1 & class 2 & best-of & SVC \\
\noalign{\smallskip}\hline\noalign{\smallskip}
0.5810 & 0.5002 & 0.4955 & 0.5045 & 0.5092 & 0.5474 \\
\noalign{\smallskip}\hline\noalign{\smallskip}
\multicolumn{6}{l}{\textbf{variances of accuracies}} \\
\noalign{\smallskip}\hline\noalign{\smallskip}
model & rand. & class 1 & class 2 & best-of & SVC\\
\noalign{\smallskip}\hline\noalign{\smallskip}
$6.12e^{-4}$ & $1.43e^{-4}$ & $5.93e^{-5}$ & $5.93e^{-5}$ & $4.50e^{-5}$ & $10.87e^{-4}$ \\
\noalign{\smallskip}\hline\noalign{\smallskip}
\multicolumn{6}{l}{\textbf{tests against baselines}} \\
\noalign{\smallskip}\hline\noalign{\smallskip}
 & rand. & class 1 & class 2 & best-of & SVC\\
\noalign{\smallskip}\hline\noalign{\smallskip}
\textit{p}-value & $< 1e^{-3}$ & $< 1e^{-3}$ & $< 1e^{-3}$ & $< 1e^{-3}$ & $< 1e^{-3}$ \\
\noalign{\smallskip}\hline\noalign{\smallskip}
min. diff. & 0.0767 & 0.0816 & 0.0727 & 0.0679 & 0.0276 \\
\noalign{\smallskip}\hline
\end{tabular}
\end{footnotesize}
\caption{Accuracies, variances and baseline comparison. Accuracies and variances for the model's predictions (\textit{model}), as well as for the best result among all baselines (\textit{best-of}), the randomized baseline with the same class distribution (\textit{rand.}), comparative results for a support vector classifier (\textit{SVC}), and results for predictions for one class exclusively are provided. \textit{class 1} denotes upward stock trend changes and \textit{class 2} downward stock trend changes. \textit{p}-values and minimum differences are provided for comparison against the chosen baselines.}
\end{table}

While the focus of this work is the existence and exploitation of lagged correlations, one might be left to wonder what impact the omission of target stock information in the inputs of each stock's run has on the results. To answer this question, we repeat the experiment described above, with this information included in the inputs. The reported result of 58.10\% without this information increases to 58.54\% when information about the target stock is included. While this shows that a stock's past behavior provides additional information, this result demonstrates the model's ability to infer most of the relevant information from lagged correlations between the target stock and other stocks.

As described in the discussion of accuracy and validation measures, a feed-forward neural network is trained on the first observation and then re-trained for 3 epochs on the dataset extended by the previous time step for the remaining time steps' prediction. This process is repeated for 20 randomly chosen stocks, after which the results for the predictions are averaged. This process is implemented until 90\% of the dataset is used as the training set in order to allow for a small number of observations to remain as a test set towards the end. Since the robustness of the obtained results is an important factor to be taken into account, leaving a sufficient amount of time steps as a test is crucial. If the process of shifting from the test to the training set would be taken to the extreme by continuing until only one stock is left in the test set, the results would be based on a non-representative number of time steps. From a financial markets perspective on robustness, the setup of this experiment allows for the assessment of the model's performance over multiple years, and thus across changing market conditions over time. For a measurement of the model's accuracy during the end of the experiments, the average of the last 100 time steps is taken and results in an average accuracy of 60.23\%.

Notably, this result outperforms the previous experiment that uses cross-validation, which indicates an advantage of learning exclusively from data before the respective test example that is to be classified instead of using training examples from both before and after the test example to extract time-invariant market dynamics. Within the framework of the stock market, an explanation for this observation is that lagged trend correlations in financial markets feature information that can be used for predictions about the future in which the lagged effect takes place, but to a lesser degree for predictions about past observations. The learned information is, in this case, local within time, rendering cross-validation less effective when compared to training on past data with online updates to adapt to new information. The accuracies of this additional experiment over the course of the training process are depicted in parts $a_1$ and $b_1$ of Figure 3.

To determine whether the initial sharp increase in average accuracy and the subsequent slower and quasi-linear increase are a result of the market or the learning, we repeat the same experiment without the first 100 time steps in which the sharp increase takes place, which means starting at a later point in time and excluding the model from learning about the training examples before that. Part $a_1$ of Figure 3 shows the initial rise in accuracy for this cut-down dataset following a steeper path at first, but reaching the full dataset's first elbow accuracy later in relation to the number of steps since starting the training process. It subsequently follows a more concave curve instead of the full dataset's quasi-linear increase in accuracy during that phase. Similarly, the final level of the cut-down dataset is reached at about the same time as the full dataset's corresponding accuracy, again in relation to the number of steps the model is trained on instead of the actual point in time due to the omission of the first 100 time steps and, therefore, a later point in real time.

The resulting average prediction accuracy is 59.50\% due to the full dataset's later and more pronounced accuracy peak, although both experiments feature the same dip around time step 900, or time step 1000 for the full dataset. While reaching the peaks in parts $a_1$ and $b_1$ of Figure 3 seems to be a result of the number of time steps the models are trained on, the dips at 900 and 1000, respectively, seem to be a result of the market at that time. These results suggests that the primary features of the time series are due to the market, whereas the profiles of the first initial accuracy increase and the subsequent rise, as well as the lower maximum towards the end, can be attributed to the lack of the information contained in the deleted first 100 training examples, inhibiting the portion of long-term market information that would otherwise be extractable from the latter.

\begin{figure}[!h]
\includegraphics[width=\columnwidth]{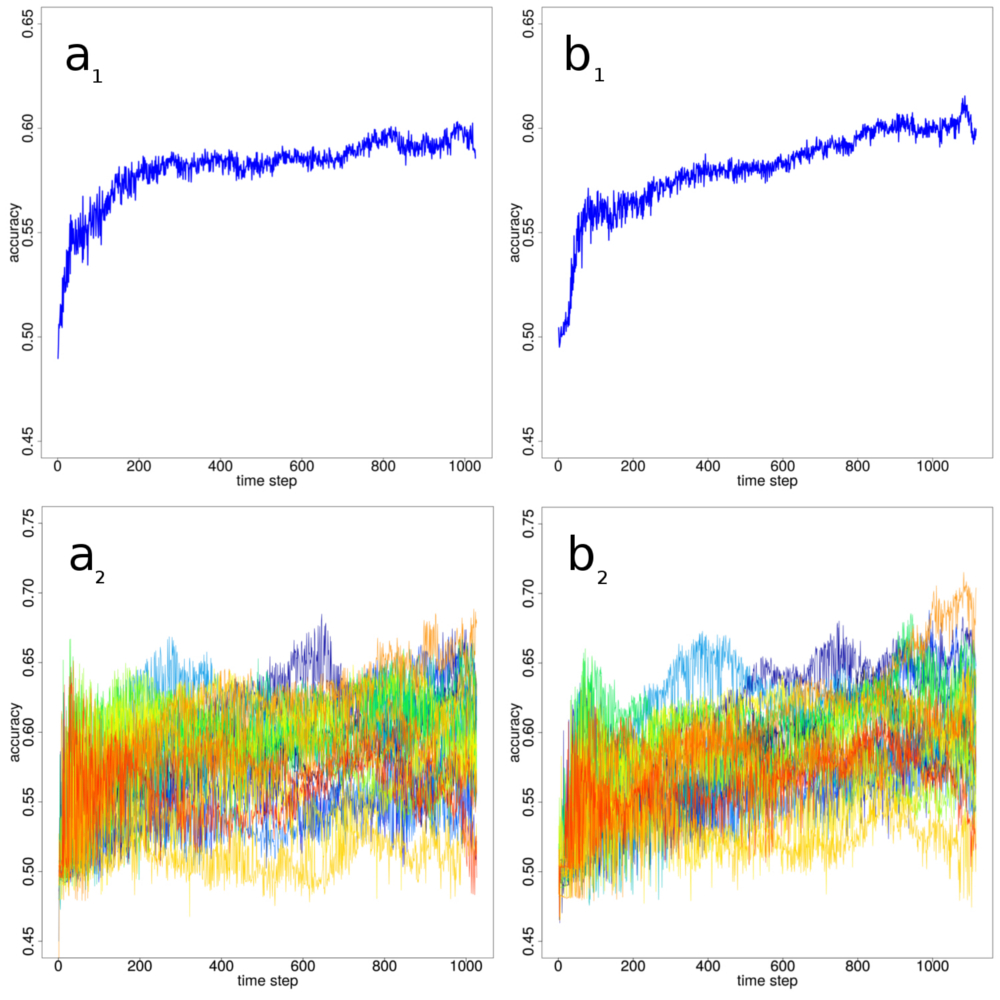}
\caption{Averaged and individual accuracies for the experiment with omitted data for the first 100 time steps. For the same 20 stocks selected via random sampling, a model is iteratively updated for 3 epochs and with the last time step's information as an extension of the training set after each new time step. Parts $a_1$ and $a_2$ show the results for averaged and individual accuracies for the case without the first 100 time steps, respectively, whereas parts $b_1$ and $b_2$ show the same visualizations for the full dataset containing the first 100 time steps.}
\label{fig:3}
\end{figure}

Another result of the omission of the first 100 time steps can be seen when comparing the individual accuracies for the targeted stocks in parts $b_1$ and $b_2$ of Figure 3. While the individual stocks in both datasets' accuracies feature similar evolutions, the full dataset's accuracies show a clearer trend towards higher accuracies. This observation reflects the more linear ascend of the average accuracy and can be viewed as an indicator of market behavior information from multiple years into the past still influencing the market behavior's overall predictability in the present, the implications of which are discussed later.

Figure 4 is comprised of heatmaps for both cases in order to get a better overview of the individual 20 stocks featured in Figure 3. As can be seen, the development of the predictive accuracy for individual stocks follows similar progressions for both experiments, albeit with slightly different distributions and starting points of high-accuracy periods. A more general finding in Figure 4 is that some stocks are considerably easier to predict solely on other stocks' prior behavior than others.

\begin{figure}[!h]
\includegraphics[width=\columnwidth]{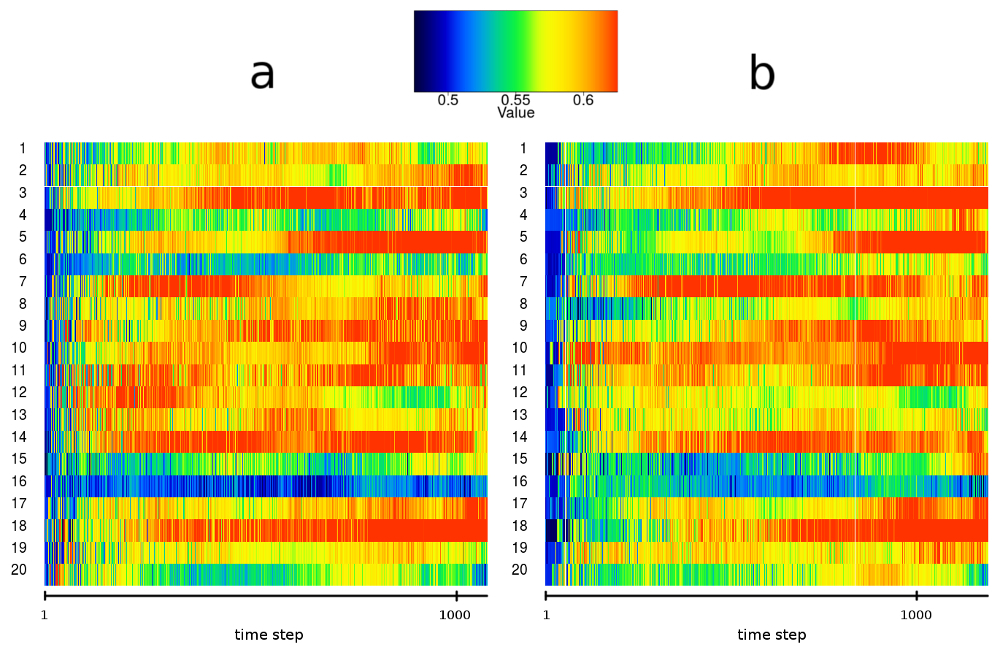}
\caption{Heatmaps for the experiment with omitted data for the first 100 time steps. For the same 20 stocks selected via random sampling, a model is iteratively updated for 3 epochs and with the last time step's information as an extension of the training set after each new time step. Part $a$ shows the results for averaged and individual accuracies for the case without the first 100 time steps, whereas part $b$ show the results for the full dataset.}
\label{fig:4}
\end{figure}

\section{Discussion}
\label{S:5}

\subsection{Relevance for trend prediction}
\label{S:5.1}

The hypothesis about general time series analysis via such network models is reinforced by the experiments: The evidence strongly suggests that deep feed-forward neural networks can be used to consistently learn and, for previously unseen data, act with an accuracy above predetermined baselines on time-shifted correlations of gradients that are computed step-wise for complex time series, with only the previous interval of other series instead of the target one as input features. While adding information about the target stock in the inputs is shown to provide only marginal increases in performance, a simple linear SVC baseline performs significantly above na\"ive baselines, further bolstering the relevance of the feature engineering used in this work. The approach of this paper could be applied to other forecasting problems that involve non-linear interactions between a large number of noisy time series and lagged effects of their respective trend behavior, for example the metrics in areas as diverse as consumer behavior and epidemic dynamics of infectious diseases. For practitioners relying on expert systems to inform decisions about the future in noise-laden environments, breaking through that noise is a common issue. As a novel use and application of gradient-based feature engineering, in combination with smoothing techniques described in Section \ref{S:3.2}, this paper delivers general evidence for the viability as an expert system in highly noisy dynamical systems subject to time series prediction problems.

\subsection{Comparison to related research}
\label{S:5.2}

Both \citet{Takeuchi2013Applying} and \citet{Batres-Estrada2015Deep} use deep learning models for the binary month-wise trend prediction of target stocks, based on historical stock market data of the preceding 12 months, with resulting accuracies of 53.36\% and 52.89\%. We want to emphasize that these papers don't work on the same time intervals and use additional features instead of just past stock prices, which means that a comparison should be taken with a grain of salt. These results are, however, the closest available comparison of feed-forward deep learning models for trend prediction based on historical stock market data. When directly comparing the accuracies, our model outperforms both by a noticeable margin, with 58.10\% for cross-validation and 60.23\% for training exclusively on past observations and re-training the model for a few iterations before each new prediction. The same, although to a lesser degree, holds true for the simple linear SVC used as an additional baseline, indicating the viability of the proposed feature engineering. 

\subsection{Implications for financial economics}
\label{S:5.3}

Firstly, the findings add to the existing evidence against the random walk hypothesis as popularized by \citet{Malkiel1973A} and others, that is, the notion that stock prices follow random walks with inherently unpredictable behavior. Our findings in this respect are consistent with a growing view in the financial economics literature. Consequently, the results also challenge the related efficient-market hypothesis developed by \citet{Fama1965The}, which has since become a staple in the field of financial economics. However, in order to convincingly argue for an empirical rebuttal of the efficient-market hypothesis, a simulation taking trading costs into account may be necessary, which could be a topic for further research and would have to show consistent outperformance in the face of these costs. Lastly, the higher accuracy for models trained exclusively on past observations and with online updates have implications for the interpretation of the underlying structures learned by the models: The latter appear to partially adapt to changing correlations between stocks, meaning that the learned information is, to a degree, temporally constrained instead of reflecting only general market behavior. This shows that stock markets are, over periods of multiple years, non-stationary in their correlated behavior even in the case of summary findings for the S\&P 500.

Another relevant finding is that the inclusion of information about the respective target stock does provide additional information that improves predictive results, but that this increase is rather small and the model is able to infer most relevant information from lagged correlations with other stocks. This recovery of most of the information relevant to the predictions in this paper further strengthens the arguments against both the random walk hypothesis and the efficient-market hypothesis in most of its forms.

The efficient-market hypothesis allows for three forms of information-based market efficiency. Our results could be consistent for a constrained version of the most basic form, the weak-form market efficiency.  Specifically, this would be for a case where the hypothesis allows for prediction methods that reliably outperform the market and can only be implemented by a sufficiently small number of investors so as not to result in a new aggregate equilibrium for a typical modern economy. With a negligible amount of capital involved in the context of the whole market, some agents, such as select quantitative hedge funds or individuals, could consistently realize above-average returns, thus reducing the weak form efficient-market hypothesis to a context-based version. A time-specific weak form efficient-market hypothesis would, in effect, acknowledge that the hypothesis does not apply to the overall market, but does for a majority of the trading stakeholders due to restrictions regarding the methodology and the capital involved in deploying such strategies. This view bears argumentative resemblance to the adaptive-market hypothesis by \citet{Lo2004The}, perceiving the efficient-market hypothesis as not necessarily incorrect, but incomplete, resulting in an attempt to merge the efficient-market hypothesis with behavioral economics by applying principles of biological evolution. For example, \citet{Lo2004The} states that a high competition for scarce resources leads to highly efficient markets, whereas a competition for abundant resources among few "species" in a financial market diminishes overall market efficiency.

\subsection{Further research suggestions}
\label{S:5.4}

High frequency trading (HFT) is the use of high-frequency stock market data characterized by short holding times and high rates of cancellation for equities and futures trading in a fully automated manner \citep{Menkveld2013High}. It remains a driving force in stock markets, with double-digit shares of total trading volumes across different markets and competition mostly between such HFT algorithms. Using our model, the interaction between those algorithms in the form of lagged correlations could be further investigated to better explain the behavior of this part of the stock market. Another approach is the use of wavelets, that is, the results of time-frequency transformation to compute a local variations representation on different scales suitable to combat noise \citep{Nason1999Wavelets}.

As described earlier, gradient-based wavelet approaches are used in the areas of natural language processing and acoustic classification, and our model could be used with wavelets as a more elaborate way to extract relevant information from intervals in time series. The question remains whether an increased sophistication equals a better model performance. While the linear SVC provides a simple off-the-shelf baseline, serving the same purpose as logistic regression in similar papers, it still performs significantly better than na\"ive baselines. As a result, further fine tuning and modification of SVCs provide a viable approach in problems where training speed is more important than accuracy maximization.

\section{Conclusion}
\label{S:6}

In conclusion, we have shown in this paper that the results with regard to the investigated hypothesis are positive under conscientious observance of statistical validation measures. The results of the experiments deliver evidence for the viability of a combination of deep feed-forward neural networks and exponential smoothing applied to gradient-based features for directional trend change predictions with non-linear correlations of large numbers of noisy time series in the form of historical stock market data. More generally, our findings demonstrate the value of deep learning approaches to time series analysis and show that linear regression derivatives provide useful features to extract such complex interdependencies, offering a simple indicator of trends with a high predictive value.

The findings in this paper also have implications for modern finance theory, delivering strong evidence against both the random walk and efficient-market hypotheses. The postulations of the efficient-market hypothesis may, however, be adapted to allow for the findings as presented here. While indeed all three forms of the efficient-market hypothesis are inconsistent with the evidence, tweaking the weak form of the efficient-market hypothesis could lead to consistency with our findings. Furthermore, while the presented findings are successfully tested on stock market data and have interesting implications for hypotheses within financial economics, they should also be applicable to other fields dealing with trend forecasts in time series. In addition, as simple arbitrage approaches to investment became less effective due to the growing use of such methods over the last decades, it remains to be seen whether deep learning methods such as the one we discuss in this paper will see a similar spread and, consequently, a failure to perform due to a large enough number of market participants operating with related techniques.


\bibliographystyle{apalike}
\bibliography{lagged_correlations.bib}

\end{document}